# Protein Dynamical Transition in Terahertz Dielectric Response

Andrea G. Markelz, Joseph R. Knab, Jing Yin Chen, and Yunfen He

Physics Department, 239 Fronczak Hall, University at Buffalo, SUNY, Buffalo, NY 14260-1500 USA

ABSTRACT

The 200 K protein dynamical transition is observed for the first time in the teraherz dielectric response. The complex dielectric permittivity $\varepsilon = \varepsilon' + i\varepsilon''$ is determined in the 0.2 - 2.0 THz and 80-294 K ranges. $\varepsilon''$ has a linear temperature dependence up to 200 K then sharply increases. The low temperature linear dependence in $\varepsilon''$ indicates anharmonicity for temperatures 80 K < T < 180 K, challenging the assumed harmonicity below 200K. The temperature dependence is consistent with beta relaxation response and shows the protein motions involved in the dynamical transition extend to subpicosecond time scales.

Address inquiries to Prof. Andrea Markelz,
239 Fronczak Hall
Physics Dept.
University at Buffalo, SUNY
Buffalo, NY 14260

Tel.: 716-645-2017 x124
Fax: 716-645-2507
Email: amarkelz@buffalo.edu

Tertiary dynamics essential for protein interaction can be inhibited by ligand binding, dehydration and temperature. The dynamics within the energy landscape [1] have been studied by the temperature dependence of protein flexibility, often examined through the average atomic mean square displacement (msd) over the molecule, $<x^2>$, determined by x-ray diffraction [2], neutron scattering (NS) intensities [3], and for iron containing proteins, Mössbauer measurements [4]. At low temperatures $<x^2>$ increases linearly and for dehydrated proteins this linear increase continues to melting temperatures. However if the samples are critically hydrated, $<x^2>$ increases rapidly near a characteristic temperature, $T_d \sim 200$ K [5] followed by in some cases a slight plateau at T ~ 250 K [5,6]. The strong temperature dependence has been labeled the dynamical transition and suggests new degrees of freedom becoming available at $T_d$ [6]. While it is widely understood that $T_d$ does not correspond to a true phase transition, (e.g. there is no heat capacity signature [7]), authors use the term "dynamical transition" to refer to *the strong dynamical increase at T ~ 200 K.* We will also use this terminology.

The linear increasing $<x^2>$ for T < 200 K agrees with harmonic vibrational response and so motions have been assumed to be harmonic in this temperature range. The mechanism for the strong $<x^2>$ temperature dependence near 200 K remains controversial (see for example [8-10]). Among the leading explanations is that motions within the protein resemble relaxational motions in glasses [8]. Specifically, after removal of the linear temperature dependence, the remaining temperature dependence is Arrenhius, suggesting glass-like beta relaxations give rise to the observed temperature dependence. The actual motions involved in these relaxations is not well defined, in part due to the lack of spectrally resolved measurements.

$<x^2>$ measurements in general have a low time cut off with integration over all faster motions. Neutron measurements generally have a cutoff of 1 ns. Mössbauer absorption measurements, sensitive to Fe coupled motions, have a cutoff of 140 ns. Earlier spectral dependent characterization using nuclear resonant vibrational spectroscopy (NRVS, a Mössbauer related technique) [11] suggested that only motions slower than 4 ps contribute to the dynamical transition.

We examine the spectral dependence of the dynamical transition in the terahertz frequency range using dielectric spectroscopy. The dielectric response of the system is related to resonant and relaxational processes through the dielectric response ε(ω):

$$\varepsilon(\omega)=\varepsilon_o+\int\frac{f(\omega')g(\omega')}{(\omega'^2-\omega^2)+i\gamma(\omega')\omega}d\omega'+\varepsilon_r\int_0^\infty\frac{h(\tau)d\tau}{1+i\omega\tau} \quad (1)$$

where $\varepsilon_o$ is the dc dielectric constant. The second term on the right hand side refers to a sum of harmonic oscillators with density of states $g(\omega)$, oscillator strength $f(\omega)$ and damping coefficient $\gamma(\omega)$. The third term refers to relaxational processes with a distribution of $h(\tau)$ Debye relaxations. We perform terahertz time domain spectroscopy (THz-TDS) as a function of temperature for cytochrome c (Cytc) solutions and find that the imaginary part of the dielectric response has a temperature dependence similar to the $<x^2>$ measurements with a linear dependence at low temperature and an Arrenhius dependence at higher temperature. However a linear temperature dependent dielectric permittivity is inconsistent with a harmonic vibrational response, and while the energy determined from Arrenhius fits (24 +/- 6 kJ/mol) is consistent with $<x^2>$ Mössbauer data [12] , the THz-TDS results indicate the motions involved in the dynamical transition extend to the subpicosecond time scale.

**Materials and Methods**

The dynamical transition observed in $<x^2>$ measurements has been shown to be hydration and solvent dependent [5]. Most THz-TDS measurements of biomolecules have been confined to hydrated/dehydrated powders or films due to the bulk water absorbance dominating the response of solutions [13-16]. This absorbance arises from $H_2O$ dipole rotation with a relaxation time of $\tau_r$ = 8.2 ps, or a peak dielectric loss at 121 GHz. $\tau_r$ scales with viscosity $\eta$ and temperature T as $\tau_r = 4\pi\eta r^3/k_bT$, where r is the molecular radius and $k_b$ is Boltzmann's constant. The high increase in bulk water viscosity with freezing leads to a shift in $H_2O$ dielectric loss to lower frequencies and the protein response dominates at these temperatures. Oxidized cytochrome C solutions are made using trizma buffer (pH 7.0) and lyophilized horse heart cytochrome c (Sigma C-2506) with a concentration of 400 mg/ml. The solution is clear with no precipitates. A brass solution cell consisting of water-free quartz windows with a 260 micron spacer is used for the transmission measurements. The top plate of the cell has two identical apertures for the sample and reference. Sufficient solution is added to fill exactly half the cell. The empty portion of the sample cell is used as a reference. A schematic of the sample cell is shown in the inset of Figure 1. The technique of THz time-domain spectroscopy, with current transient THz generation and electro optic detection, was used to probe the dielectric response [17,18]. A Ti-sapphire oscillator (82 MHz, 65 fs pulse width and 350 mW power) was used to both generate and detect the THz pulses. The entire spectroscopy system is enclosed in a plexiglass box and purged with nitrogen gas prior to and during the measurements in order to eliminate absorption due to atmospheric water rotational lines. The sample cell is loaded into a gas exchange cryostat (Cryoindustries) with liquid nitrogen used as the cryogen and low density polyethylene inner and outer windows. A transmission measurement consists of toggling between the sample and reference apertures with the toggling distance adjusted to account for any thermal contraction of the probe stick. The protein solution measurements were repeated for several fills of the cell to determine reproducibility. Measurements were also made on pure buffer solutions.

As THz-TDS measures both the transmitted field amplitude and phase, it allows for the direct extraction of the real and imaginary parts of the dielectric response. The field transmission is given by:

$$t = \frac{E_{sample}}{E_{reference}} = |t| e^{i\phi}$$
$$= e^{ik_o d[(n_s - 1) + i\kappa_s]} \frac{t_{ws} t_{sw}}{t_{wa} t_{aw}} \frac{1 - r_{aw}^2 e^{i2k_o d}}{1 - r_{sw}^2 e^{i2k_o d(n_s + i\kappa_s)}} \quad (2)$$

The first exponential factor on the right hand side is the usual propagation term through a sample with thickness d, and complex index $N_i$ related to the dielectric response $\varepsilon$ through $N_i = \sqrt{\varepsilon_i} = n_i + i\kappa_i$. The second factor is the Fresnel transmission term and the third factor arises from multiple reflections within the cell, called the Fabry-Perot term. The Fresnel transmission and reflection coefficients used in the Fresnel and Fabry-Perot terms are $t_{ij}$ and $r_{ij}$, for the ith to jth interfaces and have the form

$$t_{ij} = \frac{2N_i}{N_i + N_j} \qquad r_{ij} = \frac{N_i - N_j}{N_i + N_j} \quad (3)$$

In equation 2 the subscript s (w) stands for the sample (window), e.g. $t_{ws}$ is the Fresnel transmission from the window to sample. Our measured index for the quartz (Spectrocell, Inc.) windows is parameterized in the 0.1-2.5 THz range as $n_w = 1.92 - 0.21\exp(-5\nu^{0.79})$, where ν is the frequency in THz. In the low loss regime we neglect the imaginary part of the index in the Fresnel terms. Using Eq. 2, the measured quartz index, gasket thickness and the measured magnitude $|t|$ and phase $\phi$ we determine $n$ and $\kappa$ for both the protein solution and pure buffer with Newton Raphson fitting at each frequency. Using the $n$ and $\kappa$ extracted from the field transmission data we then determine the complex permittivity, $\varepsilon = \varepsilon' + i\varepsilon''$ with $\varepsilon' = n^2 - \kappa^2$ and $\varepsilon'' = 2n\kappa$.

**Results**

The magnitude of the electric field transmission, |t|, as a function of frequency for several temperatures is shown in Fig. 1 for the sample and pure buffer solutions. For T > 273 K the transmission of the pure buffer is smaller than the protein solution, demonstrating the significant absorbance of bulk water. The absorbance is less for the protein solution as the optically dense solvent

is partially displaced by the protein. For T < 273 K the transmission of the pure buffer is larger than the protein solution, due to the decrease in solvent absorbance with the dramatic increase in $H_2O$ rotational relaxation time. This is the first demonstration of THz-TDS measurements of fully hydrated biomolecular samples. In Fig. 1 the oscillations in |t| arise from multiple reflections within the sample cell. The greater than one transmission is due to referencing to an empty cell and the index matching of the protein solution with the cell's quartz windows. The temperature dependence |t| is shown for several frequencies in Fig. 2 for two different samples. A strong decrease in transmission occurs near 200K with a plateau at 250K.

The real and imaginary part of the index, $N = n + i\kappa$ extracted from the magnitude and phase of the field transmission are shown in Fig. 3. The etalon oscillations are not entirely removed, possibly due to inaccuracies in the sample thickness or index of the quartz. The values for the buffer are in good agreement with values determined by other groups [19,20]. The temperature dependent peak in $\kappa$ with accompanying inflection in n at 1.6 THz cannot be ascribed to etalon effects and is under further investigation as is the 0.25 THz feature in $n$.

The components of the complex permittivity as discussed in Eq. 1, $\varepsilon = \varepsilon' + i\varepsilon''$ , is plotted versus temperature for several frequencies for the protein solution along with the buffer data for 1.32 THz in Fig. 4. The trends we discuss were observed for the whole measurement range. Focusing on the 1.32 THz data, $\varepsilon''$ of the protein solution is more than twice that of the pure buffer over the full temperature range. The linear temperature dependence of the pure buffer agrees with Zhang et al's results [20]. The temperature dependence of the protein solution $\varepsilon''$ is initially linear and then rapidly increases near 200K. On the other hand $\varepsilon'$ only slightly increases above the pure buffer value above 200 K. The real part of the index, n, dominates $\varepsilon'$ and it is possible that our current sensitivity is not sufficient to see the complementary temperature dependence in $\varepsilon'$.

The temperature dependence appears similar to that $<x^2>$ as measured by Mössbauer and NS. In general the temperature dependent $<x^2>$ results have been described by the sum of a linear term and an Arrenhius contribution. We apply a similar description, that is $\varepsilon'' = \varepsilon''_\ell + \varepsilon''_A$, where $\varepsilon''_\ell$ is linear in temperature and $\varepsilon''_A$ has an Arrenhius temperature dependence $\varepsilon''_A = Ce^{-\bar{E}/k_b T}$. We will discuss the possible origins of these terms shortly. Linear fits to values of ε" for T<180 K at each frequency determine $\varepsilon''_l$, as shown by the solid line for the 0.49 THz data in Figure 3A. The linear fits are subtracted from ε" to determine $\varepsilon''_A$. The plot of $\log(\varepsilon''_A)$ versus 1000/T is shown in the inset of Fig. 3A). $\varepsilon''_A$ indeed shows Arrenhius behavior with an activation energy of 24 +/- 6 kJ/mole, consistent with the values extracted from Mössbauer results for Cytc (10).

**Discussion**

A linear temperature dependent $<x^2>$ is expected for a harmonic solid, however this is not the case for ε. In the low power regime ε is not dependent on the amplitude of the motions, but rather the VDOS and dipole coupling to the modes. For a harmonic potential ε should have no temperature dependence as the vibrational level spacing is constant and dipole coupling is temperature independent. A linearly increasing ε" can reflect a linear increase in $<x^2>$ when the potential is anharmonic. For example for a 1D potential with a quartic term:

$$V(x) = \frac{1}{2}kx^2 - \frac{1}{12}bx^4$$
$$\omega = \sqrt{\frac{\frac{\partial^2 V(x)}{\partial x^2}}{m}} \qquad (4)$$
$$\sim \omega_o (1 - b\langle x^2 \rangle_{Vib} / 2k)$$

assuming that $b/k<<1$. For this anharmonic potential a temperature dependent increase in $<x^2>$ results in a temperature dependent red shifting of the modes. This red shifting would give rise to a linear increase in the low frequency dielectric absorbance at low temperatures. We note that a recent letter using NS measurements also suggests anharmonic response below 200 K [21].

The Arrenhius fits with energy E = 24 kJ/mole are consistent with Mössbauer measurements and add confirmation to a beta relaxation source of the transition. However the THz-TDS measurements show the motions involved in the dynamical transition extend to subpicosecond frequencies, contrary to NRVS measurements [11]. How can these two contradicatory results coexist? First we note that NRVS measures the VDOS coupled to the isotopically substituted Fe, whereas the THz-TDS probes the entire system. We suggest the different results arise from two sources: higher energy motions are more localized, and the Arrenhius activation energy is hydration dependent. The data from Roh et al. shows the temperature at which there is a strong increase in $\langle x^2 \rangle$ becomes higher as the protein is dehydrated, that is the activation energy increases with decreasing hydration [22]. Beta relaxations in the protein may always be present over a broad time scale, but in the dehydrated system the activation energy is sufficiently high that the Arrenhius knee is not observable for the temperature range typically measured. As the hydration of the system is increased, the activation energy for relaxations coupled to the hydration layer lowers. For low frequency extended motions coupled to the protein surface, the solvent coupling lowers the activation energy so the transition is observed. The activation energy of higher frequency *localized* motions will only decrease with hydration if they are sufficiently coupled to the solvation shell, such as side chain motion at the surface. For localized *internal* beta motions weakly coupled to the hydration shell, such as < 1 ps motions coupled to the heme Fe, the activation energy remains high and the transition is imperceptible up to 300 K. The NRVS and THz-TDS measurements can be reconciled if the THz contribution arises from localized surface modes not strongly coupled to the heme Fe.

The source of the plateau at 250 K is unclear. The plateaus seen in <$x^2$> measurements have been explained by a maximum displacement allowed by structure [8], however we cannot use this model as THz-TDS measures VDOS and dipole coupling, not amplitudes.

In conclusion, the spectral dependence of the dynamical transition in ε" is measured using THz-TDS. A linear temperature dependence indicates anharmonicity at low temperatures and the dynamical transition at 200 K is observed up to 2.0 THz. THz-TDS allows rapid characterization of the dynamical transition without requiring large sample volumes or isotopic substitution. Further we demonstrate that THz-TDS measurements on biomolecular systems need not be limited to pressed pellets and hydrated films.

Acknowledgements. We are grateful to NSF IGERT DGE0114330, NSF CAREER PHY-0349256 and ACS 39554-AC6 for support of this work.

**References**

[1] H. Frauenfelder, S. G. Sligar, P. G. Wolynes, Science 254 (1991) 1598.

[2] D. Ringe, G. A. Petsko, Biophys. Chem. 105 (2003) 667.

[3] J. C. Smith, Quarterly Reviews of Biophysics 24 (1991) 227.

[4] F. Parak, Methods in Enzymology 127 (1986) 196.

[5] A. M. Tsai, D. A. Neumann, L. N. Bell, Biophysical Journal 79 (2000) 2728–2732.

[6] W. Doster, S. Cuszck, W. Petry, Nature 337 (1989) 754.

[7] J. L. Green, J. Fan, C. A. Angell, J. Phys. Chem. 98 (1994) 13780.

[8] P. W. Fenimore, H. Frauenfelder, B. H. McMahon, R. D. Young, Proc. Natl. Acad Sci. U.S.A. 101 (2004) 14408–14413.

[9] F. G. Parak, Rep. Prog. Phys. 66 (2003) 103–129.

[10] S.-H. Chen, L. Liu, E. Fratini, P. Baglioni, A. Faraone, E. Mamontov, Proc. Natl. Acad Sci. U.S.A. 103 (2006) 9012.

[11] K. Achterhold, C. Keppler, A. Ostermann, U. v. Bu¨rck, W. Sturhahn, E. E. Alp, F. G. Parak, Phys. Rev. E 65 (2002) 051916.

[12] E. N. Frolov, R. Gvosdev, V. I. Goldanskii, F. G. Parak, J. Biol. Inorganic Chem. 2 (1997) 710.

[13] L. Genzel, F. Kremer, A. Poglitsch, and G. Bechtold, . Biopolymers, 1983. 22(7): p. 1715-1729, Biopolymers 22 (1983) 1715.

[14] M. Brucherseifer, M. Nagel, P. H. Bolivar, H. Kurz, A. Bosserhoff, R. Buttner, Appl. Phys. Lett. 77 (2000) 4049.

[15] C. F. Zhang, E. Tarhan, A. K. Ramdas, A. M. Weiner, S. M. Durbin, Journal of Physical Chemistry B 108 (2004) 10077.

[16] A. G. Markelz, A. Roitberg, E. J. Heilweil, Chem. Phys. Lett. 320 (2000) 42

[17] D. Grischkowsky, N. Katzenellenbogen, in T.C.L. Sollner, J. Shah (Eds.), OSA Proceedings on picosecond electronics and optoelectronics. OSA, Washington, DC, 1991.

[18] Q. Wu, X.-C. Zhang, Appl. Phys. Lett. 67 (1995) 3523.

[19] J. T. Kindt, C. A. Schmuttenmaer, J. Chem. Phys. 100 (1996) 10373.

[20] C. Zhang, K.-S. Lee, X.-C. Zhang, X. Wei, Y. R. Shen, Appl. Phys. Lett. 79 (2001) 491.

[21] J. H. Roh, V. N. Novikov, R. B. Gregory, J. E. Curtis, Z. Chowdhuri, A. P. Sokolov, Phys. Rev. Lett. 95 (2005) 038101.

[22] J. H. Roh, J. E. Curtis, S. Azzam, V. N. Novikov, I. Peral, Z. Chowdhuri, R. B. Gregory, A. P. Sokolov, Biophys. J. 91 (2006) 2573–2588.

Figure Captions

Figure 1

Terahertz field transmission data for Cytc solution (—) and buffer reference (- - - ) at several temperatures. Same color coding for both types of samples.

Figure 2

Temperature dependence of magnitude of field transmission for several frequencies for two different cytochrome c samples. The plateau near 250K is reproducible, and the transmission rapidly decreases with increasing temperature near 200K.

Figure 3

The temperature dependent components of the complex index of refraction as determined from field transmission measurements. A) imaginary part of the index $\kappa$ and B) real part of the index n. 294 K values shown in insets. Same color coding for A) and B).

Figure 4

Temperature dependence of the complex permittivity $\varepsilon = \varepsilon' + i\varepsilon''$ for a Cytc solution at several frequencies and pure buffer at 1.32 THz. The inset shows $\log(\varepsilon_A'')$ vs 1/T. Same legend for A) and B).

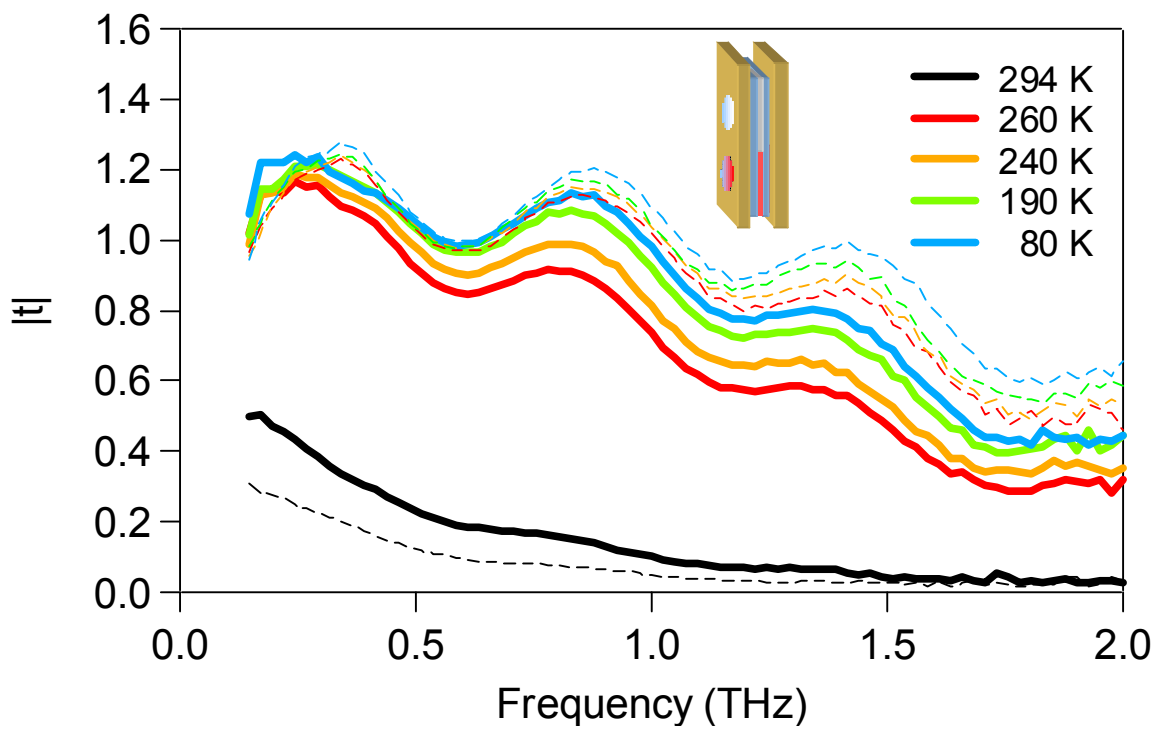

Figure 1

"Protein Dynamical Transition in Terahertz Dielectric Response,"
Markelz, Knab, Chen and He

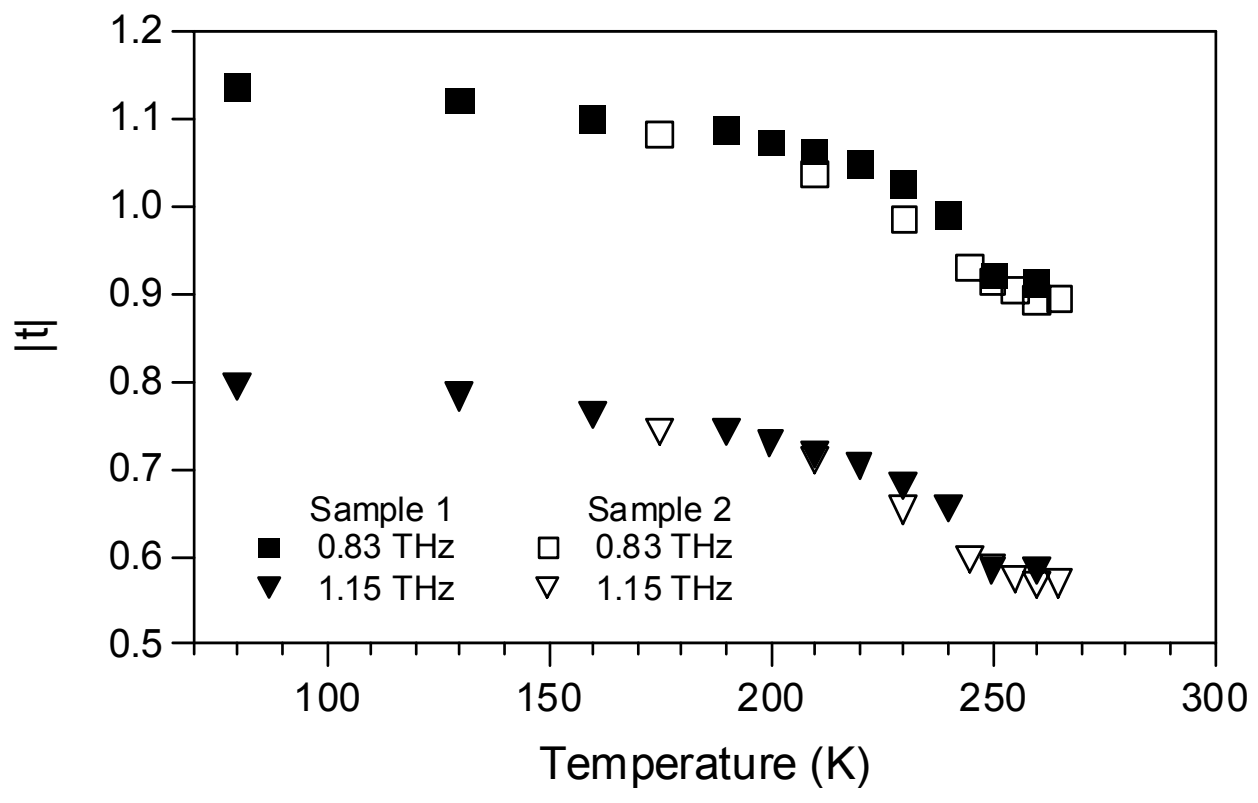

Figure 2

"Protein Dynamical Transition in Terahertz Dielectric Response,"
Markelz, Knab, Chen and He

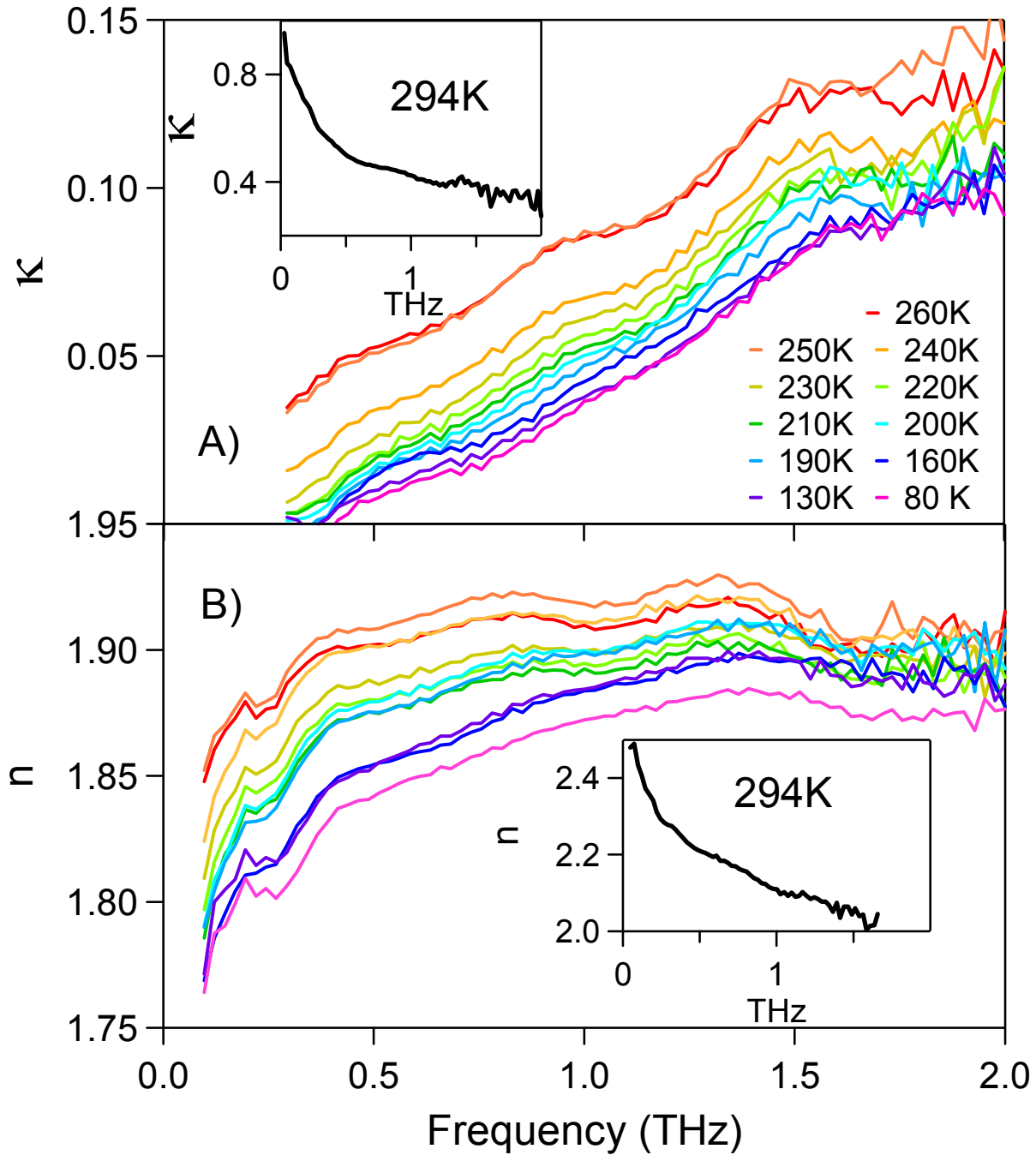

Figure 3

"Protein Dynamical Transition in Terahertz Dielectric Response,"
Markelz, Knab, Chen and He

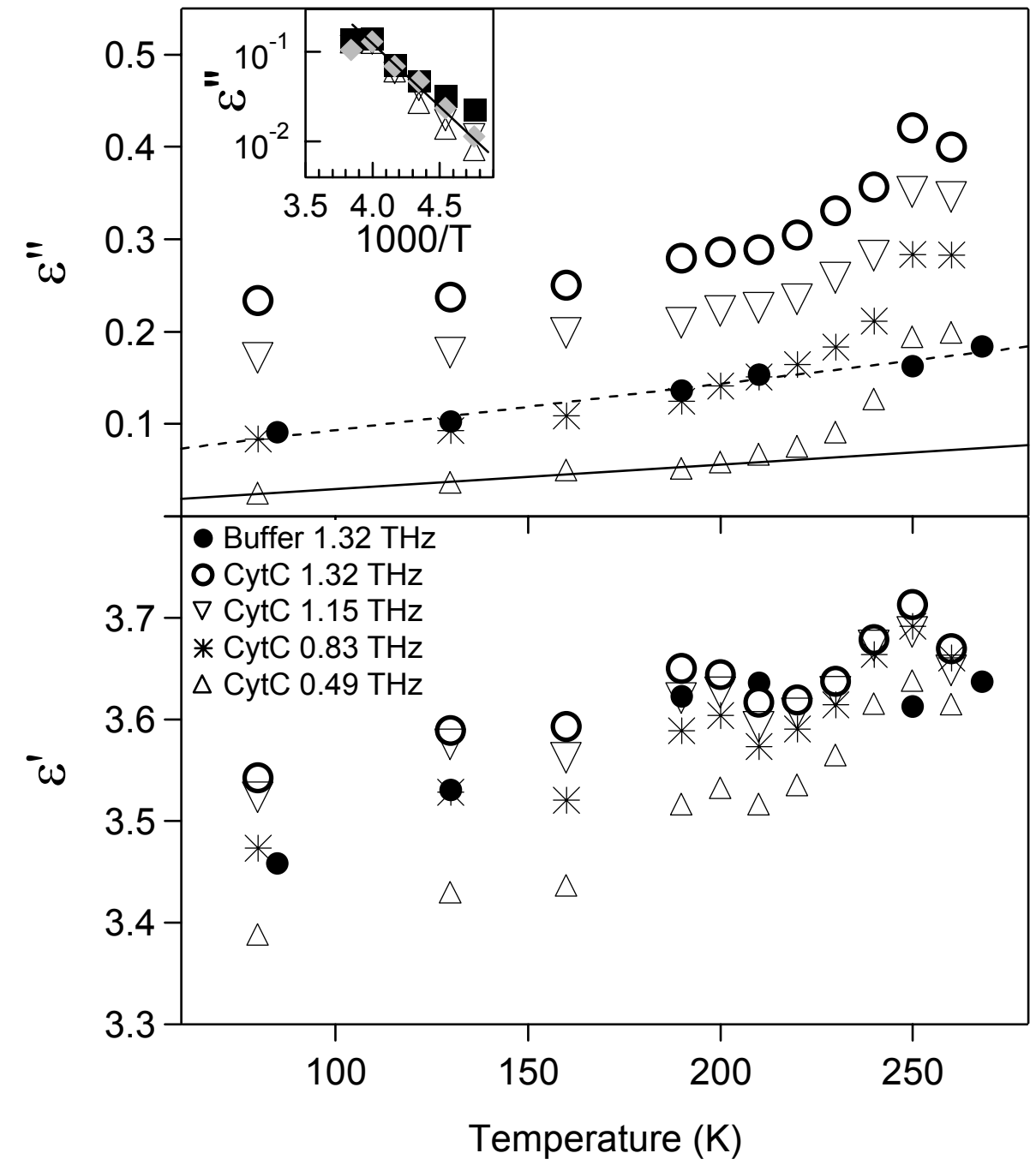

Figure 4

"Protein Dynamical Transition in Terahertz Dielectric Response,"
Markelz, Knab, Chen and He